\documentclass[12pt]{article}
\usepackage{graphicx}
\usepackage{dcolumn}
\usepackage{bm}
\usepackage{graphics}
\usepackage{amsmath}
\usepackage{amssymb}
\usepackage{amscd}
\usepackage{afterpage}
\usepackage{float,times}
\usepackage{subfigure}
\usepackage{rotating}
\usepackage{multirow}
\usepackage{epsfig}
\usepackage{theorem}
\usepackage{moreverb}
\usepackage{euscript}
\usepackage{psfrag}

\begin{document}

\begin{center}
{\hbox to\hsize{\hfill July 2008 }}

\bigskip

\vspace{6\baselineskip}

{\Large \bf

Domain walls and gauge field localization in \\ strongly-coupled pure Yang-Mills theories \\}

\bigskip

\bigskip

\bigskip

{\bf Archil Kobakhidze \\}

\smallskip

{ \small \it
School of Physics, The University of Melbourne, Victoria 3010, Australia \\
E-mail: archilk@unimelb.edu.au
\\}

\vspace*{1.0cm}

{\bf Abstract}\\
\end{center}
\noindent
{\small We present a mechanism of gauge field localization on a domain wall within the framework of 
strongly coupled pure Yang-Mills theory.
}
\bigskip

\bigskip

\baselineskip=16pt

\section{Introduction}
Some time ago, Dvali and Shifman proposed a field-theoretic mechanism for the localization of a gauge field on a hypersurface (brane) 
embedded in higher dimensional spacetime (bulk) \cite{Dvali:1996xe}. The physical picture behind this mechanism is as follows.  A gauge field is assumed to be in the confining phase in the bulk and in the Coloumb (or less confining) phase on the brane. Then the field is prevented from spreading out into extra dimensions because of the mass gap generated by the bulk confinement. Although this idea is physically rather transparent, its explicit implementation requires an adequate treatment of quantum effects in non-perturbative regime. The string theory D-brane interpretation of gauge field localization has been discussed in \cite{Dvali:1998ms} using domain wall solutions within large N supersymmetric Yang-Mills theories. An attempt to describe Dvali-Shifman mechanism based on some models of dual superconductivity has been made in \cite{Dubovsky:2001pe}. In \cite{Shifman:2002jm} more involved mechanism has been proposed within N=2 supersymmetric Yang-Mills theories which operates in weak coupling regime. 

Most of the explicit models for the Dvali-Shifman localization of gauge fields has been discussed in supersymmetric setting. Also, besides the Yang-Mills fields, the models in \cite{Dvali:1996xe} and \cite{Shifman:2002jm},  involve many additional fields and somewhat {\it ad hoc} mass parameters. Thus, although these models represent the interesting testing ground for theoretical ideas, they are far more complicated compared to a pure Yang-Mills theory where one may think that the non-perturbative mechanism  of \cite{Dvali:1996xe} could also be realized. 

In this paper we aim to demonstrate explicitly the phenomenon of gauge field localization in strongly coupled non-supersymmetric pure Yang-Mills theory. In field-theoretic approach brane is usually viewed as a topological defect (domain wall, string, ...) formed upon the condensation of a scalar field. In supersymmetric theories \cite{Dvali:1996xe},\cite{Shifman:2002jm} such scalar fields are naturally present as the constituents of supermultiplets. But in pure Yang-Mills theories there are no elementary scalar fields that could condense. However, as has been argued in \cite{Faddeev:2001dda}, \cite{Faddeev:2006sw}, effective Higgs-like fields can emerge in strongly coupled Yang-Mills theories at low energies. The physical picture in strongly coupled Yang-Mills theories is analogous to the electron spin-charge separation \cite{Anderson:1987gf} occurring in strongly correlated systems, such as in higher temperature superconducting cuprates. In what follows we will argue that condensation of the Higgs-like effective scalars support a domain wall structure on which massless Abelian gauge field gets localized.

In the present paper we explicitly consider  (3+1)-dimensional strongly-coupled SU(2) gauge theory with an Abelian gauge field localized in (2+1)-dimensions. We hope to present generalization to higher dimensions and localized non-Abelian fields in future publications. Having such an explicit field-theoretic mechanism would greatly advance more detailed studies of various interesting domain wall based higher-dimensional models  \cite{Davies:2007xr}.  

Aside the higher-dimensional theories, the physics discussed in this paper might be relevant for QCD in the infrared domain. Interestingly, a number of recent numerical studies within the lattice gauge theory framework \cite{deForcrand:2006my} as well as analytical investigations \cite{Kondo:2008xa} have also obtained evidences for the non-trivial topological structure of the nonperturbative QCD ground state. The results of our paper can be viewed as complimentary to those studies.   

\section{Spin-charge separation in SU(2) Yang-Mills}
We start with a brief review of the spin-charge separation mechanism in SU(2) gauge theory \cite{Faddeev:2001dda}.  SU(2)-valued gauge potential can be written as:
\begin{equation}
\hat A_{\mu}=A_{\mu}\tau^3+W^-_{\mu}\tau^-+W^+_{\mu}\tau^+,
\label{1}
\end{equation}  
where, $W^{\pm}_{\mu}=(A^1_{\mu}\mp iA^2_{\mu})/\sqrt{2}$, $\tau^{\pm}=(\tau^1\pm i\tau^2)/\sqrt{2}$, and $\tau^i$ are the Pauli matrices.  SU(2)Yang-Mills Lagrangian then takes the form 
\begin{eqnarray}
{\cal L}_{YM}=-\frac{1}{4}F_{\mu\nu}F^{\mu\nu}-W^{-}_{\mu}\left[\eta^{\mu\nu}D_{\alpha}D^{\alpha}
-D^{\mu}D^{\nu}-2igF^{\mu\nu}\right]W^{+}_{\nu} \nonumber \\
-\frac{1}{4}g^2\left(W_{\mu}^{+}W_{\nu}^{-}-W_{\nu}^{+}W_{\mu}^{-}\right )^2~,
\label{2}
\end{eqnarray}
where $F_{\mu\nu}=\partial_{\mu}A_{\nu}-\partial_{\nu}A_{\mu}$ and $D_{\mu}=\partial_{\mu}-igA_{\mu}$ is the covariant derivative relative to the Abelian field $A_{\mu}$ gauging the Cartan subgroup $U(1)_C$ of SU(2).

Next, the off-diagonal gauge fields $W^{+}_{\mu}=(W^{-}_{\mu})^{*}$ we decompose as
\begin{equation}
W^{+}_{\mu}=i\phi_1e_{\mu}-i\phi_2^{*}e_{\mu}^{*}~.
\label{3}
\end{equation}
In the above decomposition $\phi_{1,2}$ are the complex scalar fields which carry $U(1)_C$ charge degrees of freedom of the W-bosons, while 
\begin{equation}
e_{\mu}=\frac{1}{\sqrt{2}}(e^1_{\mu}+ie^{2}_{\mu})\equiv e^{i\sigma}\hat e_{\mu}~,
\label{4}
\end{equation}
is a complex vector with the orthonormality condition
\begin{equation}
e^i_{\mu}e^{j\mu}=\delta^{ij}~.
\label{5}
\end{equation} 
This vector describes the spin degrees of freedom of the $W$ fields. It is believed that in the non-perturbative regime at low energies precisely $\phi_i$ and $e^{i}_{\mu}$ (along with the remaining Abelian (Cartan subalgebra) field $A_{\mu}$) degrees of freedom, rather than the original gauge bosons ($A_{\mu},~ W^{\pm}_{\mu}$) of SU(2), more adequately describe the dynamics, including the anticipated spontaneous generation of the mass gap. 

Some important remarks on the above formalism are in order. First note that $\phi_i$ and $e^{i}_{\mu}$ have the same number of degrees of freedom as the original off-diagonal gauge fields,  i.e. 8. Indeed, two complex scalars $\phi_{i}$ have 4 and a complex vector $e_{\mu}$ with 3 constraints (\ref{5}) has 5 degrees of freedom, so 9 in total. However, one of these 9 degrees of freedom is redundant because of the local phase invariance,  
\begin{equation}
\phi_{1}\to {\rm e}^{i\alpha(x)}\phi_{1},~\phi_{2}\to {\rm e}^{i\alpha(x)}\phi_{2},~ e_{\mu}\to {\rm e}^{-i\alpha(x)}e_{\mu}~, 
\end{equation}
induced by the decomposition, and hence the degrees of freedom match as they should. The composite field,
\begin{equation}
B_{\mu}=ie_{\alpha}^{*}\partial_{\mu}e^{\alpha}~,
\label{a1}
\end{equation}
plays the role of the gauge boson of this $U(1)_{\alpha}$ invariance.  In \cite{Faddeev:2001dda} the field (\ref{a1}) is referred to as the magnetic gauge field and $U(1)_{\alpha}$ as the magnetic gauge invariance.    To remove the redundant degree of freedom, we instead consider $U(1)_{\alpha}$-gauge invariant transverse magnetic field, 
\begin{equation}
\hat B_{\alpha}\equiv i\hat e_{\mu}^{*}\partial_{\alpha}\hat e^{\mu}=B_{\alpha}-\partial_{\alpha}\sigma~,~~\partial^{\alpha}\hat B_{\alpha}=0~,
\end{equation}
where $\hat e_{\mu}=e^{i\sigma(x)}e_{\mu}$, and also, $U(1)_{\alpha}$-gauge invariant fields, 
$\hat\phi_1=e^{-i\sigma(x)}\phi_1$ and $\hat\phi_1=e^{-i\sigma(x)}\phi_1$. With hatted fields we have right number degrees of freedom, namely 8, and no $U(1)_{\alpha}$ local gauge invariance. However, there remains still the global phase invariance, $U(1)_{\alpha}^{\rm glob}$, 
\begin{equation}
\hat\phi_{1}\to {\rm e}^{i\alpha}\hat\phi_{1},~\hat\phi_{2}\to {\rm e}^{i\alpha}\hat \phi_{2},~ \hat e_{\mu}\to {\rm e}^{-i\alpha}\hat e_{\mu}~. 
\end{equation}
This will be important in what follows.

Another important point is that the decomposition in (\ref{3}) is not gauge invariant in general and thus physical relevance of various quantities requires special care. We are interested in gauge invariant observables, and in particular we would like to have a gauge invariant order parameter,
\begin{equation}
v^2=\left\langle |\phi_1|^2 \right\rangle + \left\langle |\phi_2|^2\right\rangle
\label{vev}
\end{equation}     
This is not invariant under the full SU(2) gauge transformations unless we impose the gauge fixing condition
\begin{equation}
D^{\mu}W^{+}_{\mu}=0~,
\label{gauge}
\end{equation}
which we assume to hold in what follows. By fixing the gauge (\ref{gauge}) SU(2) symmetry is reduced and the remaining local gauge invariance becomes $U(1)_C$,  where $U(1)_C$ is the diagonal Cartan subgroup of the original SU(2), under which the fields transform as
\begin{equation}
\hat\phi_1\to {\rm e}^{i\theta(x)/2}\hat\phi_1,~ 
  \hat\phi_2\to {\rm e}^{-i\theta(x)/2}\hat\phi_2~, \hat e_{\mu}\to \hat e_{\mu}~.
\end{equation}
Thus the full symmetry group is $U(1)_C\times U(1)^{\rm glob}_{\alpha}$.

Finally, we stress non-canonical behavior of some composite objects introduced through the spin-charge decomposition (\ref{3}) under the Lorentz transformations. Namely, as is discussed in detail in \cite{Faddeev:2006sw}, while the object $\frac{|\hat \phi_1|^2-|\hat \phi_2|^2}{|\hat \phi_1|^2+|\hat \phi_2|^2}\equiv\cos2\beta$ transforms as a Lorentz scalar (it is also $U(1)_C \times U(1)_{\alpha}$-invariant), the objects $\frac{2\hat \phi_1\hat \phi_2}{|\hat \phi_1|^2+|\hat \phi_2|^2}$ and $\frac{2\hat \phi_1^{*}\hat \phi_2^{*}}{|\hat \phi_1|^2+|\hat \phi_2|^2}$ are realized as projective representations.\footnote{The complex vector field $\hat e_{\mu}$  also transforms as a projective representation, so that the Lorentz invariance of the total Lagrangian is maintained.} This will be important in the next section, where we will discuss the ground state. Indeed, due to the constraint, 
\begin{equation}
\cos^22\beta+\frac{4|\hat \phi|_1^2|\hat \phi_2|^2}{(|\hat \phi_1|^2+|\hat \phi_2|^2)^2}=1, 
\label{aa}
\end{equation}
non-trivial Lorentz invariant vacuum state can be realized if, and only if,
\begin{equation}
\langle \cos2\beta \rangle =\pm 1~. 
\label{a2}
\end{equation}

Summarizing this section, we have expressed original SU(2) Yang-Mills theory with gauge bosons $(A_{\mu}, W_{\mu}^{+}, W_{\mu}^{-})$ in terms of new variables $(A_{\mu}, \hat B_{\mu}, \hat \phi_1, \hat\phi_2)$. In the gauge (\ref{gauge}) the Lagrangian (\ref{2}) takes the form:
\begin{eqnarray}
{\cal L}_{YM}=-\frac{1}{4}F_{\mu\nu}F^{\mu\nu}+\left |(\partial_{\mu}-igA_{\mu}-i\hat B_{\mu} )\hat \phi_1\right |^2+
\left |(\partial_{\mu}+igA_{\mu}-i\hat B_{\mu} )\hat \phi_2\right |^2  \nonumber \\
+(|\hat \phi_1|^2+|\hat \phi_2|^2)(\partial_{\alpha}\hat e_{\mu}^{*}\partial^{\alpha}\hat e^{\mu}-\hat B_{\alpha}\hat B^{\alpha})+(\hat \phi_1\hat \phi_2\partial_{\alpha}\hat e_{\mu}\partial^{\alpha}\hat e^{\mu}+ {\rm c.c.})+4igF^{\mu\nu}\hat e_{\mu}\hat e^{*}_{\nu}(|\hat \phi_1|^2-|\hat \phi_2|^2) \nonumber \\
-\frac{1}{8}g^2\left(|\hat \phi_1|^2-|\hat \phi_2|^2 \right)^2~.
\label{a3}
\end{eqnarray}

\section{The ground state}
With the aim to determine the ground state of the theory given by (\ref{a3}), let us concentrate on the last non-derivative term in (\ref{a3}), 
\begin{equation}
V_0(\phi_1, \phi_2)=\frac{1}{8}g^2\left(|\hat \phi_1|^2-|\hat \phi_2|^2 \right)^2
\label{6}
\end{equation} 
\begin{figure}
\centering
\includegraphics[width=0.4\textwidth]{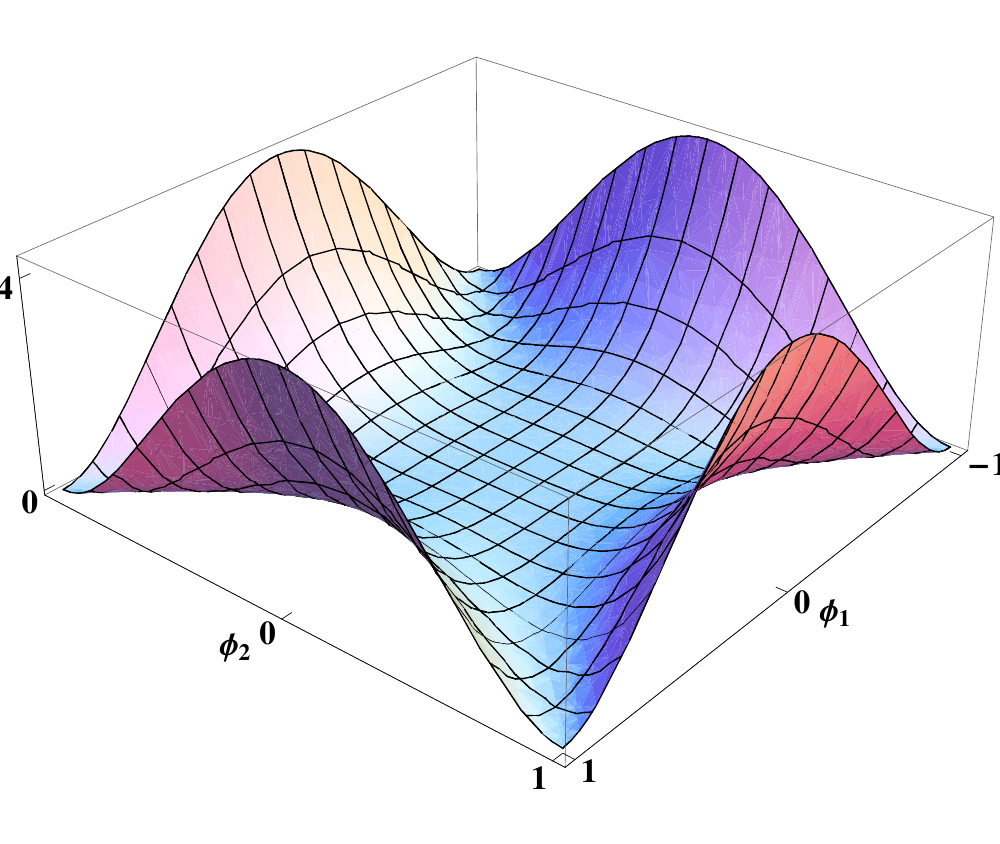}
\qquad
\includegraphics[width=0.4\textwidth]{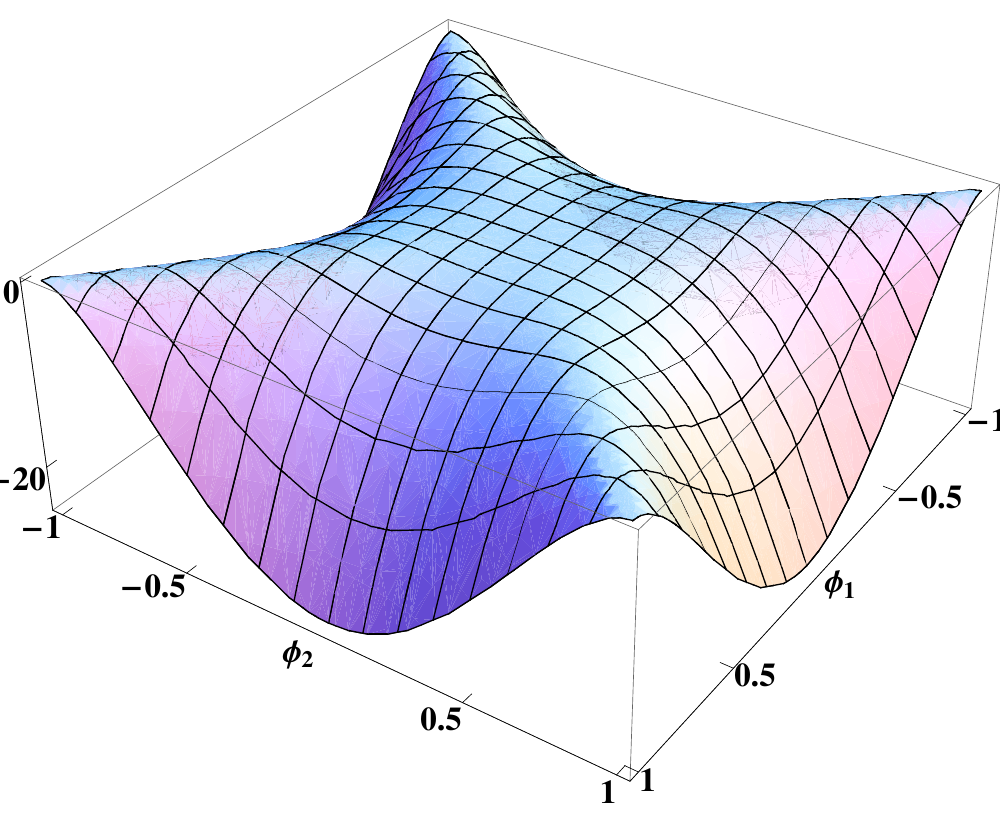}
  \caption{\small 3D plots of the tree level $V_{0}$ (on the left) and one-loop $V_1$ (on the right) potentials as a functions of $(\hat \phi_1, \hat \phi_2)$. Mass is given in units of $v$. The Lorentz-invariant vacuum for $V_0$ corresponds to trivial $(0,0)$ configuration, while at one-loop level one has non-trivial vacua $(\pm 1, 0)$ and $(0, \pm 1)$. The non-trivial vacua, e.g. $(1,0)$ and $(0,1)$, are separated by the potential barrier.  }
  \label{figure1}
\end{figure}
This potential is shown in Figure 1 (the left graph).  The Lorentz-invariant ground state is realized by homogeneous field configurations $\left\langle|\hat \phi_{1,2}|\right \rangle=v_{1,2}$ subject to the constraint (\ref{a2}).  It is easy to see that at tree level the ground state is given by the trivial configuration, $v_1=v_2=0$.  In order to locate the true ground state one must inspect radiative corrections to (\ref{6}). This has been done in \cite{Freyhult:2001ds} (see also \cite{Niemi:2005qs}) and we just borrow the 1-loop effective potential obtained there:
\begin{equation}
V_1=V_0(v_1,v_2)-\beta_{{\rm SU(2)}}(g^2)\frac{(v_1^2-v_2^2)^2}{4} \left(
\log \frac{\sqrt{|v_1^2-v_2^2|}}{\mu}-\frac{25}{12}~.
\right)
\label{7}
\end{equation}
$\beta_{{\rm SU(2)}}(g^2)=-\frac{22}{3}\frac{g^4}{16\pi^2}$ is the 1-loop SU(2) $\beta$-function for the running gauge coupling $g(\mu)$: $\frac{dg}{d\log{\mu}}=\frac{\beta_{{\rm SU(2)}}}{g}$. The potential (\ref{7}) is drawn in Figure 1 (the right graph). The renormalization scale is convenient to associate with the gauge invariant order parameter (\ref{vev}), $\mu=v$ The minima now satisfy the equation, 
\begin{equation}
\frac{|v_1^2-v_2^2|}{v_1^2+v_2^2}\equiv \langle |\cos2\beta| \rangle =\exp\left( -\frac{6\pi}{11\alpha}+\frac{11}{3}\right)~,
\label{min}
\end{equation}
where $\alpha=\frac{g^2}{4\pi}$ is the SU(2) fine structure constant. We observe from (\ref{min}) that nontrivial Lorentz-invariant solutions occur for 
\begin{equation}
\alpha(v)=\frac{18\pi}{121}\approx 0.47. 
\label{b}
\end{equation}
This value is on the edge of perturbation theory where the 1-loop results still can be trusted.\footnote{Obviously, the simple 1-loop approximation cannot give quantitatively accurate predictions \cite{Leutwyler:1980ma}. Nevertheless, we believe that the qualitative picture of the vacuum structure is adequately described at 1-loop level.}  The equation (\ref{b}) in principle determines the vacuum expectation value $v$ in terms of dimensionless SU(2) gauge coupling defined at certain energy scale, the phenomenon known as the dimensional transmutation.

Notice the constraint (\ref{aa}) induces global O(3) symmetry. However the Lorentz-invariant ground state has no O(3) degeneracy, but is only doubly degenerate under discrete exchange symmetry, $v_1\leftrightarrow v_2$ , since the Lorentz invariance strictly implies (\ref{a2}). Therefore, we have
\begin{equation}
v_1\equiv v\langle \cos\beta \rangle=v,~~ v_{2}\equiv v\langle \sin\beta \rangle=0~, \\
\label{mina}
\end{equation}
\begin{equation}
v_1\equiv v\langle \cos\beta \rangle=0,~~ v_{2}\equiv v\langle \sin\beta \rangle=v~.  
\label{minb}
\end{equation}
Thus a solution for fields $\hat \phi_1=f_1(z)$, $\hat \phi_2=f_2(z)$  interpolating, e.g., between  vacua  (\ref{mina}) at $z=-\infty$ and (\ref{minb}) at $z=+\infty$ will describe topologically stable domain wall\footnote{We assume for definiteness that the wall lies in the $xy$ plane and is centered at $z=0$} (see also \cite{Rozowsky:2003qr} for a similar domain-wall configuration).  This, the so-called "standard" domain wall configuration, is obtained numerically and is depicted in Figure 2.   

\begin{figure}
\centering
\includegraphics[width=0.9\textwidth]{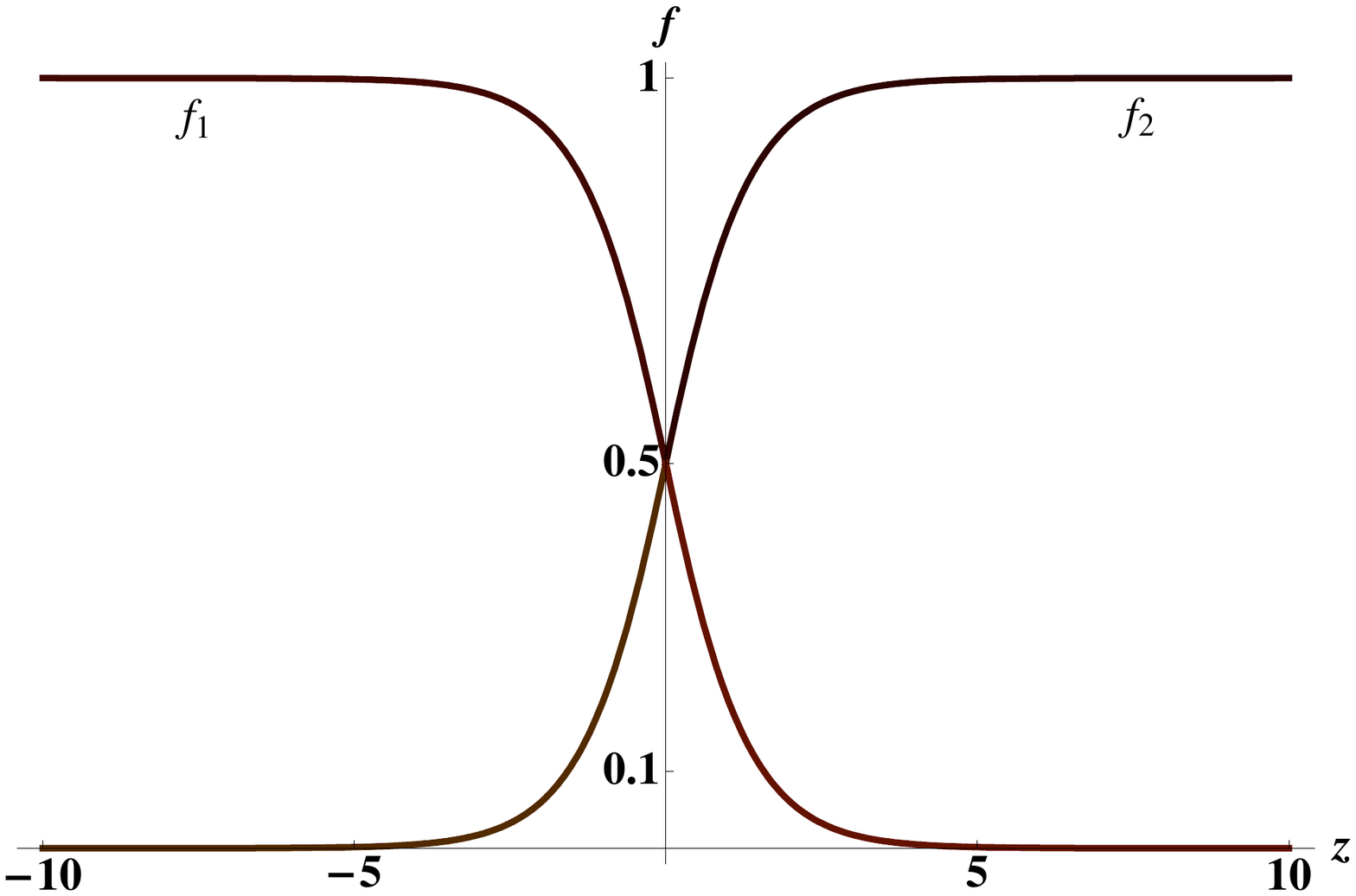}
\vspace{-2cm}
  \caption{\small Plot of the numerical solution for the "standard" domain wall corresponding to the field configuration interpolating between vacua $(1,0)$ and $(0,1)$. Mass is given in $v$ units.}
  \label{fig}
\end{figure}

\section{Gauge field on the wall} 

The above domain-wall configuration is not the most general one. In fact a family of solutions can be obtained by multiplying the "standard" domain wall configuration by a phase,  
\begin{equation}
F_{1,2}=f_{1,2}(z)e^{i\sigma/2}~. 
\label{p}
\end{equation} 
Indeed, one of the two $\hat \phi$ fields, always can be taken to be real due to the  $U(1)_C$ gauge invariance (this is just the gauge fixing condition). However, the (asymptotic) solutions on either side of the standard domain wall are continuously degenerated due to the $U(1)_{\alpha}^{\rm glob}$ global invariance which is spontaneously broken {\it only} on the domain wall. Thus  $\sigma$ in (\ref{p}) represents a collective coordinate of the wall which is realized as the Goldstone boson of spontaneously broken  $U(1)_{\alpha}^{\rm glob}$, $\sigma=\sigma(t,x,y)$.\footnote{The physical picture here fully parallels the one discussed in \cite{Shifman:2002jm}.}   The low-energy Lagrangian for this mode looks as \cite{Dvali:1998ms}, \cite{Shifman:2002jm}: 
\begin{equation}
{\cal L}_{\sigma}=\frac{1}{2\ell g^2}\partial_m\sigma\partial^{m}\sigma~,
\end{equation}  
where $m=0,1,2$ and $\ell \sim (gv)^{-1}$. On the other hand, compact scalar field $\sigma$ is dual to the (2+1)-dimensional gauge field, 
\begin{equation}
{\cal F}_{mn}=i\epsilon_{mnk}\partial^{k}\sigma~.
\label{j}
\end{equation}
Then if we identify (2+1)-dimensional gauge coupling $g^2_{(2+1)}=\ell g^2$ we obtain the following Lagrangian, 
\begin{equation}
{\cal L}=-\frac{1}{4g_{(2+1)}^2}{\cal F}_{mn}{\cal F}^{mn}~, 
\end{equation}
which describes (2+1)-dimensional Abelian gauge field localized on the domain wall. 

One can envisage that the above mechanism for gauge field localization is in accord with the heuristic argument suggested in \cite{Dvali:1996xe} and represents field-theoretic analogue ( see \cite{Dvali:1998ms} and \cite{Shifman:2002jm} ) of the D-brane picture in string theory. Far away the domain wall $v\neq 0$, and the magnetic field $\hat B_{\mu}\equiv (B_{\mu}-\partial_{\mu}\sigma)$ becomes massive.  The spontaneous breaking of the local $U(1)_{\alpha}$ in the bulk must also result in the formation of magnetic Abrikosov-Nielsen-Olesen strings. The fluctuation of the string end-points attached to the wall gives the mode $\sigma (t,x,y)$ which in turn is dual to the electric field ${\cal F}_{mn}$ according to the relation (\ref{j}). It would be interesting to explore this picture in more quantitative details.   

\section{Conclusion}

In this paper we have found the domain wall solution in strongly coupled pure SU(2) Yang-Mills theory based on the spin-charge decomposition formalism developed in \cite{Faddeev:2001dda}, \cite{Faddeev:2006sw}. The domain wall supports localized collective coordinate of the wall which is dual to the Abelian gauge field. Thus we have found that the mechanism for gauge field localization discussed in \cite{Dvali:1996xe}, \cite{Shifman:2002jm}  can be naturally realized in pure Yang-Mills theories.

\paragraph{Acknowledgments.}
I am indebted to Ray Volkas  for useful discussion,  and to Nadine Pesor for her suggestion on numerical calculations.  This work was supported by the Australian Research Council.



\end{document}